\documentclass[twocolumn,pra,showpacs,amsmath,amssymb,nofootinbib]{revtex4}
\usepackage{amsthm,url}

\begin{document}
\title{Reply To ``Comment on `Quantum String Seal Is Insecure' ''}
\author{H. F. Chau}
\email{hfchau@hkusua.hku.hk}
\affiliation{Department of Physics, University of Hong Kong, Pokfulam Road,
 Hong Kong}
\affiliation{Center of Theoretical and Computational Physics, University of
 Hong Kong, Pokfulam Road, Hong Kong}
\date{\today}

\begin{abstract}
 In Phys. Rev. A {\bf 76}, 056301 (2007), He claimed that the proof in my
 earlier paper [Phys. Rev. A {\bf 75}, 012327 (2007)] is insufficient to
 conclude the insecurity of all quantum string seals because my measurement
 strategy cannot obtain non-trivial information on the sealed string and escape
 detection at the same time.  Here, I clarify that our disagreement comes from
 our adoption of two different criteria on the minimum amount of information a
 quantum string seal can reveal to members of the public.  I also point out
 that He did not follow my measurement strategy correctly.
\end{abstract}
\maketitle
 In Ref.~\cite{comment}, He said that the key measurement strategy used in my
 paper~\cite{stringsealinsecurity} to show the insecurity of all quantum string
 seals was in the form 
\begin{equation}
 Q_{i0} = a(\nu) I + b(\nu) |i\rangle\langle i| ~. \label{E:Q_i0}
\end{equation}
 (See Eq.~(29) in Ref.~\cite{stringsealinsecurity}.)  He then attempted to find
 a ``loophole'' in my conclusion by means of two ``counter-example'' quantum
 string seals.

 His first scheme (Scheme~A) is a family of quantum string seals each with a
 different sealed string length.  In Scheme~A, each bit in a string is
 independently encoded as a publicly accessible state in the form $\cos
 \theta_i |b_i\rangle + \sin \theta_i |\bar{b}_i\rangle$, where $b_i$ is the
 value of the bit and $| \theta_i | \leq \Theta / n^\alpha$.  Here $\Theta \ll
 \pi / 4$ and $\alpha < 1/2$ are two positive constants, and $n$ is the bit
 string length.  The probabilities of correctly determining a particular bit of
 the string and the entire string are $\approx \cos^2 ( \Theta / n^\alpha )$
 and $\approx \cos^{2n} ( \Theta / n^\alpha )$, respectively
 \cite{comment,hestringseal}.  In other words, as $n\rightarrow\infty$, the
 chance of correctly finding out the entire string is negligible even though
 the percentage of correctly determined bits approaches $1$.  Furthermore, the
 number of incorrectly determined bits increases without bound in the large $n$
 limit.  In his second scheme (Scheme~B), the classical message $i$ is encoded
 as a publicly accessible quantum state $\sum_j \lambda_{ij} |j\rangle$.  He
 claimed that (the magnitudes of) $\lambda_{ij}$'s should be very small (large)
 if the contents of the messages $i$ and $j$ were irrelevant (close)
 \cite{comment}.  In both schemes, He said that to follow the instructions in
 my paper \cite{stringsealinsecurity}, a member of the public (Bob) had to
 apply the measurement operators in the form $Q_{i0}$'s to each encoded qubit.
 Nevertheless, this measurement strategy has little chance to obtain
 non-trivial information on the sealed string and escape detection
 simultaneously.  He further claimed that Bob is extremely likely to be caught
 when the parameter $\nu$ used in $Q_{i0}$ approaches $1$ because the maximum
 probability of correctly determining the entire sealed string can be made
 arbitrarily small by increasing $n$ \cite{comment}.  One can judge the
 validity of He's claim by answering the following two questions: What is the
 minimum (classical) information a quantum string seal can reveal to Bob?  And
 is He really using my measurement strategy reported in
 Ref.~\cite{stringsealinsecurity} in both schemes?

 To answer the first question, let us recall that the objective of a quantum
 seal is to allow detection of Bob's measurement with a high probability
 without concealing the sealed message \cite{originalbitseal}.  There are at
 least three possible ways to define what is the meaning of non-concealment in
 this context; and I list them in the order of decreasing ability to recover
 the sealed message.

\begin{description}
 \item[Criterion~A.] There is a measurement for Bob in such a way that the
  conditional entropy $H_\text{cond}$ of the sealed message given the
  measurement results is less than a fixed non-negative number $H_\text{crit}$
  independent of the string length $n$.  Moreover, the mutual information
  ${\cal I}$ between the sealed message and the measurement result divided by
  the entropy $H$ of the sealed message is of the order of $1$.  Thus the
  expected number of bits whose values are wrongly determined by Bob is finite
  even though Bob's chance of correctly determining the entire sealed string
  can be low.  \label{Itm:1}
 \item[Criterion~B.] The value $H_\text{cond} / H \rightarrow 0$ (and hence
  ${\cal I} / H \rightarrow 1$) in the limit of $n\rightarrow\infty$.  In other
  words, the percentage of incorrectly determined bits approaches $0$ in the
  large $n$ limit although the number of incorrectly determined bits may
  approach infinity.  \label{Itm:2}
 \item[Criterion~C.] The value $H_\text{cond} / H < c$ (and hence ${\cal I} /
  H \geq 1-c$ in the large $n$ limit, where $c$ is a fixed positive number of
  order of $1$.  That is, the percentage of incorrectly determined bits is
  bounded by a non-zero value in the large $n$ limit.  \label{Itm:3}
\end{description}

 Clearly, Scheme~A proposed by He satisfies Criterion~B but not Criterion~A.
 Furthermore, independently sealing each bit of a classical string by an
 imperfect quantum bit seal is an example of a family of quantum seals obeying
 only Criterion~C.

 A major source of the disagreement between He and myself comes from the fact
 that I have adopted Criterion~A while He used the more lenient Criterion~B.  I
 believe that it is more natural to adopt the non-concealment Criterion~A as
 the expected Hamming distance between the Bob's measurement result and the
 sealed message is bounded.  Since most of the discussion in
 Ref.~\cite{stringsealinsecurity} was focused on sealing a fixed finite number
 of possible messages $N$, the distinction between the above three possible
 non-concealment criteria was not clearly made there.

 Now I answer the second question: Is He using my measurement strategy reported
 in Ref.~\cite{stringsealinsecurity} in his analysis?  As I have already
 pointed out in the second page of Ref.~\cite{stringsealinsecurity} that the
 maximum probability for Bob to correctly determine the entire sealed string
 can be small.  Thus the first step in constructing the measurement strategy
 for a quantum string seal is to find a partition ${\mathfrak P}$ of the set of
 all possible sealed messages so that the maximum probability for Bob to
 correctly determine which element in the partition does the sealed message
 belong to $p_\text{max}$ is of the order of $1$.  In the case of Scheme~A, a
 possible choice is to partition the $N = 2^n$ possible values of the sealed
 bit string according to the values of its first $n^{2\alpha}$ bits. In the
 large $n$ limit, $p_\textrm{max}$ for this choice equals $\exp ( - \Theta^2 )
 > 0.5$.  In addition, such a probability can be attained by measuring the
 first $n^{2\alpha}$ qubits in the standard basis and keeping the remaining
 qubits untouched; and contrary to He's claim of using $Q_{i0}$'s, my
 measurement strategy reported in Ref.~\cite{stringsealinsecurity} is to apply
 the measurement operators $M_i (\nu)$'s defined by Eqs.~(11) and~(12) in
 Ref.~\cite{stringsealinsecurity}.  It is straight-forward to check that using
 these measurement operators, Bob can obtain $n^{2\alpha}$ bits of information
 on the sealed string and escape detection with at least $0.5^2 = 0.25$ chance
 simultaneously for any $1/2 \leq \nu \leq 1$.  That is, as $n\rightarrow
 \infty$, the amount of information on the sealed message obtained is infinite
 although the percentage of information obtained is $0$.  A comparison of my
 measurement strategy on families of quantum seals satisfying the three
 different non-concealment criteria are tabulated in Table~\ref{T:1}.  In
 particular, only by adopting Criterion~A is it always possible to find a
 partition ${\mathfrak P}$ satisfying $\log |{\mathfrak P}| \lesssim \log N$
 and $p_\text{max}$ is of the order of $1$.

\begin{table}[t]
 \begin{tabular}{|c|c|c|c|c|}
  \hline
  Criterion & $H_\text{cond}$ & $\displaystyle \frac{H_\text{cond}}{H}$ &
   $\log | {\mathfrak P} |$\footnote{Hence also the scaling of the average
    number of bits that can be extracted without being caught.} &
   $\displaystyle \frac{\log | {\mathfrak P} |}{n}$\footnote{Hence also the
    scaling of the number of bits that can be extracted without being caught
    divided by the string length.} \\
  \hline
  A & $< H_\text{crit}$ & 0 & \multicolumn{1}{|c|}{\parbox{2.8cm}{$+\infty$,
   scales linearly with $n$}} &
   1 \\
  & & & & \\
  B & $\leq +\infty$ & 0 &
   \multicolumn{1}{|c|}{\parbox{2.8cm}{$+\infty$, but may scale sublinearly
    with $n$}} &
   \multicolumn{1}{|c|}{\parbox{2cm}{can be $0$ in the worst case}} \\
  & & & & \\
  C & $\leq +\infty$ & $\geq 0$ &
   \multicolumn{1}{|c|}{\parbox{2.8cm}{can be finite in the worst case}} &
   \multicolumn{1}{|c|}{\parbox{2cm}{can be $0$ in the worst case}} \\
  \hline
 \end{tabular}
 \caption{Properties of families of quantum seals satisfying different
  non-concealment criteria in the limit $n\rightarrow\infty$.}
 \label{T:1}
\end{table}

 Let me further clarify.  Measurement operators $M_i$'s can be applied to
 \emph{any} quantum string seal to obtain information with a high chance of
 escaping detection.  In contrast, $Q_{i0}$'s in the form of Eq.~(\ref{E:Q_i0})
 are the measurement operators that maximize the average fidelity between the
 measured state and the sealed state for a special quantum string seal designed
 to prove Theorem~1 in Ref.~\cite{stringsealinsecurity} only.  In fact, for a
 general quantum string seal, the measurement operators that maximize the
 average fidelity between the measured state and the sealed state need not be
 in the form $Q_{i0}$'s.  One example is the (perfect) quantum string seal that
 encodes the classical message $i$ as $|\phi_i\rangle \equiv \sum_{j=0}^{N-1}
 \omega_N^{i j} |j\rangle / \sqrt{N}$ for all $i=0,1,\ldots ,N-1$, where
 $\omega_N$ is a primitive $N$th root of unity.  Clearly, one can correctly
 obtain the entire sealed message without being caught using the projective
 measurement operators $|\phi_i\rangle\langle\phi_i |$'s
 \cite{bechmannbitseal,bitsealinsecurity}.  Besides, these operators are not in
 the form of Eq.~(\ref{E:Q_i0}).  This example also shows that, contrary to
 He's claim in the security analysis of Scheme~B \cite{comment}, it is possible
 that the magnitudes of $\lambda_{ij}$'s are all equal for all $i,j$.

 In conclusion, it is the adoption of different non-concealment criteria that
 causes the disagreement between He and myself on the security of quantum
 string seals.  Moreover, the issue is further complicated by He's misuse of
 $Q_{i0}$'s as the measurement operator in his security analysis.  As shown in
 Table~\ref{T:1}, all quantum string seals are insecure if the non-concealment
 Criterion~A is used.  However, some quantum string seals are secure if
 non-concealment Criteria~B or~C is used in the large $n$ limit.  Note further
 that in the case of adopting Criterion~C, even an honest Bob can obtain only a
 fraction of the sealed message in the large $n$ limit.

 Finally, I remark on passing that He's claim in Ref.~\cite{comment} that by
 linearity, the operator $\alpha I + \beta |i\rangle\langle i|$ actually meant
 applying the identity operator with a certain probability is incorrect.  This
 is because the state $\alpha |\psi\rangle + \beta \langle i | \psi\rangle
 |i\rangle$ does not equal a mixture of pure states $|\psi\rangle$ and
 $|i\rangle$.

\begin{acknowledgments}
 This work was supported by the RGC grant No.~HKU~7010/04P of the HKSAR
 Government.
\end{acknowledgments}

\bibliography{qc38.2}
\end{document}